\documentclass[11pt,epsf]{article}

\usepackage{epsfig}
\usepackage{amssymb}
\usepackage{graphicx}
\usepackage{color}

\begin{document}

\baselineskip 6mm
\renewcommand{\thefootnote}{\fnsymbol{footnote}}


\newcommand{\nc}{\newcommand}
\newcommand{\rnc}{\renewcommand}



\newcommand{\tcb}{\textcolor{blue}}
\newcommand{\tcr}{\textcolor{red}}
\newcommand{\tcg}{\textcolor{green}}


\def\be{\begin{equation}}
\def\ee{\end{equation}}
\def\ba{\begin{array}}
\def\ea{\end{array}}
\def\bea{\begin{eqnarray}}
\def\eea{\end{eqnarray}}
\def\nn{\nonumber\\}


\def\ct{\cite}
\def\la{\label}
\def\eq#1{(\ref{#1})}


\def\a{\alpha}
\def\b{\beta}
\def\g{\gamma}
\def\G{\Gamma}
\def\d{\delta}
\def\D{\Delta}
\def\e{\epsilon}
\def\et{\eta}
\def\ph{\phi}
\def\Ph{\Phi}
\def\ps{\psi}
\def\Ps{\Psi}
\def\k{\kappa}
\def\l{\lambda}
\def\L{\Lambda}
\def\m{\mu}
\def\n{\nu}
\def\th{\theta}
\def\Th{\Theta}
\def\r{\rho}
\def\s{\sigma}
\def\S{\Sigma}
\def\ta{\tau}
\def\o{\omega}
\def\O{\Omega}
\def\pr{\prime}


\def\half{\frac{1}{2}}

\def\goto{\rightarrow}

\def\na{\nabla}
\def\grad{\nabla}
\def\curl{\nabla\times}
\def\div{\nabla\cdot}
\def\pa{\partial}

\def\bra{\left\langle}
\def\ket{\right\rangle}
\def\lb{\left[}
\def\lc{\left\{}
\def\ls{\left(}
\def\lp{\left.}
\def\rp{\right.}
\def\rb{\right]}
\def\rc{\right\}}
\def\rs{\right)}
\def\fr{\frac}

\def\vac#1{\mid #1 \rangle}


\def\td#1{\tilde{#1}}
\def\check{ \maltese {\bf Check!}}


\def\Tr{{\rm Tr}\,}
\def\det{{\rm det}}


\def\bc#1{\nnindent {\bf $\bullet$ #1} \\ }
\def\ch {$<Check!>$ }
\def\ss {\vspace{1.5cm}}

\begin{titlepage}

\hfill\parbox{5cm} { }

\vspace{25mm}

\begin{center}
{\Large \bf Nucleon mass splitting in the isospin medium}

\vskip 1. cm
  {Bum-Hoon Lee$^{a,b}$\footnote{e-mail : bhl@sogang.ac.kr},
  Shahin Mamedov$^{a,c}$\footnote{e-mail : sh.mamedov62@gmail.com} , and
  Chanyong Park$^d$\footnote{e-mail : cyong21@ewha.ac.kr} }

\vskip 0.5cm

{\it $^a\,$Center for Quantum Spacetime (CQUeST), Sogang University, Seoul 121-742, Korea}\\
{ \it $^b\,$Department of Physics, Sogang University, Seoul 121-742, Korea}\\
{ \it $^c\,$Institute for Physical Problems, Baku State University, Z.Khalilov 23, Baku, AZ-1148, Azerbaijan}\\
{\it $^d\,$Institute for the Early Universe, Ewha Womans University, Daehyun 11-1, Seoul 120-750, Korea}
\end{center}

\thispagestyle{empty}

\vskip2cm


\centerline{\bf ABSTRACT} \vskip 4mm

\vspace{1cm}

Using the AdS/CFT correspondence, we investigate a nucleon mass splitting and pion-nucleon coupling
in the isospin medium. We find that there exists a nucleon mass splitting
which is exactly given by the half
of the meson mass splitting because nucleon has the half isospin charge  of
the charged mesons. In addition, we also investigate the pion-nucleon 
and four pion interactions which
requires the modification of the known Abelian-type unitary gauge fixing 
because the non-Abelian structure of the isospin medium.
After finding the non-Abelian unitary gauge fixing,
we find that in the isospin medium the couplings only for $\pi^0 \pi^0 \pi^+ \pi^-$ and $\pi^0 
\pi^0  \pi^0 \pi^0 $ of four pion interactions shift,
while there is no pion-nucleon coupling splitting
in spite of the nucleon's and meson's mass splittings.

\vspace{2cm}


\end{titlepage}

\renewcommand{\thefootnote}{\arabic{footnote}}
\setcounter{footnote}{0}



\section{Introduction}

Understanding strongly interacting systems is one of big issues in physics. However,
there is no traditional systematic tool to explain such phenomena. Recently, there was
a new challenging idea to understand strongly interacting systems like the low energy QCD
and condensed matter system by using the holographic method, the so called AdS/CFT
correspondence
\cite{Maldacena:1997re,Gubser:1998bc,Witten:1998qj,Aharony:1999ti,Sachdev:2010ch}. In this paper, we will apply
the holographic method to the low energy QCD in the isospin medium and investigate the nucleon
mass spectra and their coupling with pion.
In order to understand physics in a certain medium, in general, it is important
to know the interaction between the medium and fundamental excitations.
In the low energy QCD, the nuclear medium is strongly
interacting system, so there is no way to explain its physics with a traditional field theory technique.
To understand such a system by using the AdS/CFT correspondence,
one first knows the corresponding dual gravity theory
and its geometric solution. There are several known geometric solutions. The thermal AdS geometry with
a hard or soft wall is dual to the confining phase of the pure gauge theory while the Schwarzschild AdS black
brane is mapped to the deconfining phase at finite temperature \cite{Erlich:2005qh}-\cite{Kim:2007em}.
These works were further generalized to the gauge theory with matter \cite{Domokos:2007kt}-\cite{Park:2014gja}.
For example, the Reissner-Nordstr\"{o}m AdS black brane can describe a
quark medium at finite temperature and its zero temperature version, the so called thermal charged AdS
geometry with a hard wall, is dual to the nuclear matter due to the confining behavior.
There is another interesting geometry representing the isospin medium, which has only the isospin
chemical potential but not the isospin number density \cite{Son:2000xc}-\cite{Albrecht:2010eg}.
Due to its simplicity and analyticity, studying the isospin medium seems to be helpful in
understanding the medium effect of the holographic QCD and condensed matter system.

There were many interesting works related to the holographic nucleon masses and nucleon-meson coupling
in the AdS space dual to the pure Yang-Mills theory \cite{Hong:2006ta}-\cite{Zhang:2010bn}.
In this case, since there is
no isospin interaction, the isospin charges of the excitations
are not important. In other words, it is sufficient to turn only on the Abelian fluctuations without
regarding
the isospin structure. This is not true anymore in the isospin medium. Since the isospin medium
has a nontrivial isospin chemical potential, charged excitations usually interacts with the
isospin medium which can change some physical properties of the excitations. For example, in the
holographic pure Yang-Mills theory dual to the AdS space proton and neutron are indistinguishable
because of the absence of the isospin interaction. So their masses become degenerate.
On the contrary, charged mesons and nucleons in the isospin or nuclear medium usually interact with the background
medium. Therefore,
one should regard the non-Abelian fluctuations to represent such isospin interactions. For the meson case,
it was shown that the isospin interaction causes the meson mass splitting \cite{Lee:2013oya}.
One can also easily expect that similar mass splitting  occurs in
nucleon  due to the nonzero isospin charges of nucleons \cite{Chang:2007sr}. 
In this paper, we will investigate holographically the isospin dependence of
the four pion interaction, the nucleon masses and pion-nucleon coupling in the isospin medium. 
In this procedure the important and crucial
thing is to find an appropriate gauge fixing which can decouple pion from the axial vector 
fluctuations.
In the isospin medium, due to the non-Abelian nature of fluctuations, the unitary gauge fixing
terms used in the AdS case \cite{Da Rold:2005zs,Hong:2006ta} should be modified. After the
careful calculation, we find that the generalized non-Abelian unitary gauge fixing is allowed in the
isospin medium which makes us be able to study the pion-nucleon coupling even in the
isospin medium. In the non-Abelian unitary gauge, we find that there exists a nucleon mass splitting.
The mass shift of nucleon is exactly given by the half of the meson mass splitting because
the isospin charge of nucleon is  half of the charged meson. Although the isospin interaction shifts meson's
and nucleon's masses, it does not affect on their Fourier mode solutions.
Using this fact we show that the couplings only for $\pi^0 \pi^0 \pi^+ \pi^-$ and $\pi^0 
\pi^0  \pi^0 \pi^0 $ are modified in the isospin medium
and that there is no pion-nucleon coupling splitting.

The rest of the paper is organized as follows: In Sec. 2, we summarize the dual geometry
and find the non-Abelian unitary gauge fixing terms in the isospin medium. In this non-Abelian
unitary gauge, the meson mass splitting are reinvestigated which 
gives rise to the consistent results obtained in the axial gauge \cite{Lee:2013oya}. 
By using the non-Abelian unitary gauge, in Sec. 3 we take into account the four pion interactions
in the isospin medium. Furthermore, we investigate
the nucleon mass splitting in Sec. 4 and the pion-nucleon couplings in Sec. 5. 
We conclude our work with some remarks in Sec. 6.

\section{Isospin medium in hard wall model}

Isospin medium is composed of matter
having only the isospin chemical potential without any baryonic net charge. To describe the
isospin medium holographically, we should take into account a gravity theory including $SU(2)_L
\times SU(2)_R$ flavor gauge group \cite{Park:2011zp,Lee:2013oya}
\begin{equation}
S=  \int d^{5}x\sqrt{G}  \ \ \Tr \left[ \frac{1}{2\kappa ^{2}}\left( \mathcal{R}-2\Lambda \right)
- \frac{1}{4g_5^{2}}  {F}^{1}_{MN} {F}^{1MN} - \frac{1}{4g_5^{2}}  {F}^{2}_{MN} {F}^{2MN}
- \left| D \Ph \right|^2 + \fr{3}{R^2} \left| \Ph \right|^2 \right] ,
\label{2}
\end{equation}%
where $\Lambda =-6/R^{2}$ is a cosmological constant. The gauge field strength are given by
\bea
F^{1}_{MN} &=& \partial _{M} L_{N} -\partial _{N} L_{M} - i\lb L_M, L_N\rb ,\nn
F^{2}_{MN} &=& \partial _{M} R_{N} -\partial _{N} R_{M} - i\lb R_M, R_N\rb , \label{5}
\eea
and a covariant derivative acting on the complex scalar field  is defined as
\be
D_M \Ph = \pa_M \Ph - i L_M \Ph + i \Ph R_M .
\ee
Since the isospin matter can be clarified by the Cartan subalgebra of the  flavor group, it is 
sufficient to turn only on $L^{3}_t$ and $R^{3}_t$  where $3$ and $t$ denote the gauge index and time component
respectively. Note that the isospin matter is not an ordinary nucleon matter, because it does not contain any information
for the baryon number.
Since the isospin chemical potential is given by a constant on the AdS background,   there is no gravitational backreaction caused by the isospin matter.
So the dual geometry still remains the thermal AdS space for the confining phase or the Schwarzschild AdS black brane for the deconfining phase.
To describe the confinement in the bottom-up approach, one should introduce a hard or soft
wall. In this paper, we will concentrate on the hard wall model \cite{Erlich:2005qh,Lee:2013oya}.
In the confining phase, the dual geometry is with an appropriate IR cutoff $z_{IR}$
\be
ds^2 = \fr{R^2}{z^2} \ls \et_{\m \n} dx^{\m} dx^{\n}  - d z^2 \rs ,
\ee
where $\et_{\m \n} = (+1, -1,-1,-1)$.

The modulus of the complex scalar field $\ph$ on the AdS background geometry satisfies the following equation of motion
\be
0 = - \fr{1}{\sqrt{G}} \pa_z \ls \sqrt{G} G^{zz} \pa_z \ph \rs + 3 \ph ,
\ee
where we set $R=1$. Its solution of $\ph$ is given by
\be		\la{res:scalarmodulus}
\ph = m_q\ z + \s \ z^3 .
\ee
In the dual theory point of view, $m_q$ and $\s$ correspond to the quark mass and the chiral condensate
respectively. In general, the energy-momentum tensor of the scalar field can modify the background
geometry. In this paper, however, we neglect the gravitational backreaction of the scalar field because
it corresponds to the $1/N_c$ correction where $N_c$ is the rank of the gauge group 
\cite{Lee:2010dh,Park:2011ab}.

The isospin matter can be described by two diagonal subgroup elements of the flavor group
and it can be further decomposed into the symmetric or antisymmetric combinations
\be
V^3_t =   \half \ls L^3_t + R^3_t  \rs \quad , \qquad A^3_t = \half \ls L^3_t - R^3_t \rs .
\ee
Under the parity transformation, $L \leftrightarrow R$, the parity-even state
usually has lower energy
than the parity-odd one. So we assume that all isospin matters are in the lowest parity-even
state satisfying $L^3_t = R^3_t$, which corresponds to the ground state of the isospin medium.
In the nuclear medium \cite{Lee:2013oya}, the boundary value of $V^3_t$ corresponds
to the difference of the isospin chemical potential.

In the isospin medium, $\m_u$ and $\m_d$ imply the isospin chemical potential of quark-like particles
without baryon charge. In order to distinguish such particles from ordinary quarks and nucleons,
we use a new terminology, iso-particle. Taking the analogy to the nuclear matter, the isospin matter implies a medium composed
of iso-nucleons which have three iso-quarks inside similar to ordinary nucleons.
Since the fundamental excitations are not quarks but nucleons in the
confining phase, we need to reinterpret $V^3_t$ in terms of iso-nucleon quantities.
Although the isospin
matter is not the nuclear matter as mentioned previously, one can still use the  same definition
 used in the nuclear
matter because the isospin matter can be regarded as the specific limit of the nuclear matter (see the details in \cite{Lee:2013oya}).
Then, $V^3_t$ can be rewritten in terms of the chemical potentials of iso-proton and iso-neutron
\bea
V^3_t = \sqrt{2} \pi^2 \ls \m_P - \m_N \rs ,
\eea
where $\m_P = 2 \m_u + \m_d$ and $\m_N= \m_u + 2 \m_d$.

To investigate the spectra and coupling of nucleons and meson, let us turn on the small 
fluctuations
\bea
L_M &\equiv& L_M^a  T^a = \ls V_M^a + l_M^a \rs T^a,\nn
R_M &\equiv& R_M^a  T^a =  \ls V_M^a + r_M^a \rs T^a ,\nn
\Ph &=& \ph \ e^{i P} = \ph \ e^{i P^a  T^a} ,
\eea
where $ l_M^a$,  $r_M^a$ and $P^a$ are fluctuations and $T^a=\sigma^a /2$ are generators of
the $SU(2)$ flavor symmetry group.
In the axial gauge $v_z=a_z=0$, the meson mass splitting in the isospin and nuclear
medium has been investigated in \cite{Kim:2007xi,Lee:2013oya}.
In \cite{Da Rold:2005zs}, the meson masses by taking the unitary gauge instead of the axial
gauge were investigated which has been further generalized to the nucleon masses
and their couplings \cite{Hong:2006ta}.
In this paper, we will investigate the nucleon masses and coupling in
the isospin medium. To do so, it is required to define meson spectra and pion
in the isospin medium correctly in the unitary gauge.
Note that in the axial gauge $v_z=a_z=0$ \cite{Erlich:2005qh,Lee:2013oya}, $P$ was
identified with the pion field.
On the other hand, authors of \cite{Da Rold:2005zs,Hong:2006ta} have introduced gauge-fixing terms
without taking $a_z=0$ and chosen the unitary gauge ($\xi_{v,a} \to \infty$). In the latter case,
the nonzero $a_z$ was identified with pion.
Here, we will follow the same strategy used in the unitary gauge.
In the isospin medium, due to the background gauge field $V_t^3$
and non-Abelian fluctuations, the unitary gauge fixing term
should be modified. To find the correct gauge fixing term in the isospin medium,
we first expand the action to quadratic order. After some calculations together with
the following redefinition
\be		\la{def:vectoraxial}
v_M = \fr{1}{\sqrt{2}} \ls l_M + r_M \rs \quad {\rm and } \quad
a_M = \fr{1}{\sqrt{2}} \ls l_M - r_M \rs ,
\ee
the action at quadratic order can be rewritten as
\be
S = S_v + S_a ,
\ee
with
\bea
S_v &=& - \fr{1}{2 g_5^2} \int d^5 x \ \sqrt{G} \ \lb \fr{}{} G^{\m\n} G^{\r\s} F^{(v)i}_{\m\r}  F^{(v)i}_{\n\s}  + 2 \ls {\cal D}_{\m} v_{z}^i  \rs^2
- 4 G^{\m\n} G^{zz}  \ls \pa_z v_{\m}^i \rs \ls {\cal D}_{\n} v_{z}^i \rs  \rp \nn
&& \qquad  \qquad  \qquad  \qquad \lp
+ 2 \ls \pa_z v_{\m}^i \rs^2   \fr{}{} \rb ,
\eea
\bea		\la{act:axial}
S_a &=& \int d^5 x \ \sqrt{G} \ \lb  - \fr{1}{2 g_5^2}  \lc G^{\m\n} G^{\r\s} F^{(a)i}_{\m\r}  F^{(a)i}_{\n\s}
+ 2  \ls {\cal D}_{\m} a_{z}^i  \rs^2 - 4 G^{\m\n} G^{zz}  \ls \pa_z a_{\m}^i \rs \ls {\cal D}_{\n} a_{z}^i \rs  + 2  \ls \pa_z a_{\m}^i \rs^2 \rc  \rp  \nn
&& \lp  \qquad \qquad \quad  \ \ + \fr{\ph^2}{2}  \ls {\cal D}_{\m} P^i - \sqrt{2} a_{\m}^i \rs^2
 + \fr{\ph^2}{2}  \ls {\pa}_{z} P^i - \sqrt{2} a_{z}^i \rs^2  \rb ,
\eea
where $i=1,2,3$ are the flavor group indices. Since we consider the non-Abelian fluctuations, the
covariant derivative ${\cal D}_{\m}$ instead of a normal derivative generally acts on fluctuations, for example,
\bea
{\cal D}_{M} \ P^i  &=& \pa_{M}  \ P^i   + \e^{i3j} \ V_M^3 \ P^j , \nn
F^{(v)i}_{MN}  &=&  {\cal D}_{M} \ v_{N}^i  - {\cal D}_{N} \ v_{M}^i , \nn
F^{(a)i}_{MN}  &=&  {\cal D}_{M} \ a_{N}^i  - {\cal D}_{N} \ a_{M}^i ,
\eea
where $\e^{ijk}$ is the structure constant of the $SU(2)$ flavor group. Note that ${\cal D}_{z}=\pa_z$
because of $V_z^i=0$. The fluctuation action includes several mixing terms which can be removed
by choosing an appropriate gauge fixing.
In \cite{Da Rold:2005zs,Hong:2006ta}, the Abelian-type gauge fixing terms in the gluon
medium without matter have been found. In the isospin medium, due to the isospin interaction,
we should turned on the non-Abelian fluctuations for which we need to generalize
the Abelian gauge fixing to the non-Abelian one. 
We find that the following non-Abelian gauge fixing terms can get rid of
most of mixing terms
\bea		\la{res:gaugefixing}
S_{GF} &=&  \int d^5 x \ \lb - \fr{1}{2 g_5^2 z \xi_v} \lc {\cal D}^{\m} v^i_{\m}
-  \xi_v z \pa_z \ls \fr{ v_z^i}{z} \rs \rc^2 \rp \nn
&& \lp \qquad \qquad - \fr{1}{2 g_5^2 z \xi_a} \lc {\cal D}^{\m} a^i_{\m}
- \xi_a z \pa_z \ls \fr{ a_z^i}{z} \rs + \xi_a
\fr{2 \sqrt{2} g_5^2 \ph^2}{z^2}  P^i  \rc^2 \rb .
\eea
Note that this non-Abelian generalization is possible
because of the constant gauge potential $V^3_t$.
In the unitary gauge ($\xi_{v,a} \to \infty$), $v_z^i$ is decoupled from the theory. Furthermore, if the following relation is satisfied
\be		\la{res:unitarresult}
P^i =  \fr{z^3}{2 \sqrt{2} \  g_5^2 \ph^2} \pa_z \ls \fr{a_z^i}{z} \rs   ,
\ee
the orthogonal combination of $a_z$ and $P$ remains massless similar to
\cite{Da Rold:2005zs,Hong:2006ta}.

Now, let us take into account the meson spectra in the non-Abelian unitary gauge in which
the action for the vector and axial-vector fluctuations simply reduce to
\bea
S_v &=& -\fr{1}{2 g_5^2} \int d^5 x \sqrt{G} \ls G^{\m\n} G^{\r\s} F^{v(i)}_{\m\r}
F^{v(i)}_{\n\s} + 2 G^{zz} G^{\m\n} \pa_z v^i_{\m} \pa_z v^i_{\n}  \rs, \nn
S_a &=& -\fr{1}{2 g_5^2} \int d^5 x \sqrt{G} \ls G^{\m\n} G^{\r\s} F^{a(i)}_{\m\r}
F^{a(i)}_{\n\s} + 2 G^{zz} G^{\m\n} \pa_z a^i_{\m} \pa_z a^i_{\n} - 4 g_5^2 \ph^2
G^{\m\n} a^i_{\m} a^i_{\n}\rs,
\eea
The nonzero value of the background gauge field in the isospin medium
breaks the boost symmetry so that the time component fluctuations usually behaves differently
from the spatial components. Moreover, since the time component fluctuations are associated
with the isospin charge rather than the meson spectra, we turn off the time component fluctuations, $v_0^i=a_0^i=0$, from now on and focus only on the spatial components.
In order to identify the bulk fluctuations with mesons of the dual field theory, we introduce the
$\r$- and $a_1$-meson notations
\bea
v^1_{\m} (z,\vec{x}) = \fr{1}{\sqrt{2}} \ls \r^+_{\m} + \r^-_{\m} \rs \ , \quad
& v^2_{\m} (z,\vec{x}) = \fr{i}{\sqrt{2}} \ls \r^+_{\m} - \r^-_{\m} \rs \ , \quad
& v^3_{\m} (z,\vec{x})= \r^0_{\m} ,  \nn
a^1_{\m} (z,\vec{x})= \fr{1}{\sqrt{2}} \ls a_{1\m}^+ + a_{1\m}^-\rs \ , \quad
& a^2_{\m}(z,\vec{x}) = \fr{i}{\sqrt{2}} \ls a_{1\m}^+ - a_{1\m}^-\rs \ , \quad
& a^3_{\m} (z,\vec{x}) = a_{1\m}^0, \quad
\eea
where the superscripts, $\pm$ and $0$, imply the isospin charges of mesons.
Using the Fourier mode expansions in the rest frame
\bea
\r_{\n}^0 (z,\vec{x}) &=& \int \fr{d \o_0}{2 \pi} \ e^{i \o_0 t} \ \r^0_{\n} (z,\o_0) , \nn
\r_{\n}^{\pm} (z,\vec{x}) &=& \int \fr{d \o_0}{2 \pi} \ e^{i \o_{\pm} t}
\ \r^{\pm}_{\n} (z,\o_{\pm}) ,
\eea
$\r$-mesons satisfy
\bea
0 &=& \pa_z \ls \sqrt{G} G^{zz} G^{\m\n} \pa_z \r^0_{\n} \rs
- \o_0^2 \sqrt{G} G^{00} G^{\m\n}  \r^0_{\n}  , \nn
0 &=& \pa_z \ls \sqrt{G} G^{zz} G^{\m\n} \pa_z \r^{\pm}_{\n} \rs
- \ls \o_{\pm} \mp V_t^3 \rs^2  \sqrt{G} G^{00} G^{\m\n}  \r^{\pm}_{\n} ,
\eea
where $\r^0_{\n}$ and $\r^{\pm}_{\n}$ are Fourier modes.
To solve these second order differential equations, one should impose two boundary conditions:
one is the Dirichlet boundary condition at the asymptotic boundary, $\lp \r^0_{\n} \right|_{z=0} =
\lp \r^{\pm}_{\n} \right|_{z=0} =0$, and the other is the Neumann boundary condition at the
IR cutoff, $\lp \pa_z \r^0_{\n} \right|_{z=z_{IR}} = \lp \pa_z \r^{\pm}_{\n} \right|_{z=z_{IR}} =0$.
Since $\r^0_{\n}$  and $\r^{\pm}_{\n}$ satisfy the same boundary conditions and $V_t^3$
is a constant, the comparison of two differential equations shows that the charged $\r$-meson masses
are related to the neutral one
\be
\o_{\pm} = \o_{0}  \pm   \sqrt{2} \pi^2 \ls \m_P - \m_N \rs .
\ee
This result is consistent with those in \cite{Lee:2013oya} where the axial gauge was used. 
Inserting the mass relation into the above differential equations, one can easily see that
$\r^{\pm}_{\n}$ should be the same as $\r^0_{\n}$.
Similarly, the Fourier mode expansions of $a_1$-meson
\bea
a_{1\n}^0 (z,\vec{x}) &=& \int \fr{d \bar{\o}_0}{2 \pi} \ e^{i \bar{\o}_0 t}
\ a_{1\n}^0 (z,\bar{\o}_0) , \nn
a_{1\n}^{\pm} (z,\vec{x}) &=& \int \fr{d \bar{\o}_0}{2 \pi} \ e^{i \bar{\o}_{\pm} t}
\ a_{1\n}^{\pm} (z,\bar{\o}_{\pm}) ,
\eea
lead to
\bea
0 &=& \pa_z \ls \sqrt{G} G^{zz} G^{\m\n} \pa_z a^0_{1\n} \rs
- \ls G^{00}  \bar{\o}_0^2 - 2 g_5^2 \ph^2 \rs \sqrt{G} G^{\m\n}  a^0_{1\n}  , \nn
0 &=& \pa_z \ls \sqrt{G} G^{zz} G^{\m\n} \pa_z a^{\pm}_{1\n} \rs
- \ls G^{00}  \ls \bar{\o}_{\pm} \mp V_t^3 \rs^2 - 2 g_5^2 \ph^2 \rs \sqrt{G} G^{\m\n}  a^{\pm}_{1\n} .
\eea
The same boundary conditions used in the $\r$-meson  give rise to the mass relation of $a_1$-mesons
\be
\bar{\o}_{\pm} = \bar{\o}_{0}  \pm   \sqrt{2} \pi^2 \ls \m_P - \m_N \rs .
\ee
Like $\r$-meson, the same boundary conditions yields $a^{\pm}_{1\n} = a^0_{1\n}$.

The action for the pseudoscalar field in the unitary gauge is given by
\be		\la{act:pseudoscalar}
S_s = \int d^5 x \sqrt{G} \lb  -\fr{1}{2 g_5^2}  G^{zz} {\cal D}_{\m} a^i_z  {\cal D}^{\m} a^i_z
+ \fr{\ph^2}{2}   {\cal D}_{\m} P^i  {\cal D}^{\m} P^i  + \fr{\ph^2}{2} \ls \pa_z P^i -  \sqrt{2} a_z^i \rs^2 \rb .
\ee
If the pion field  is identified with a component of $a_z$
\be		\la{res:modespion}
a_z^i (x,z) = f_0^i (z)  \ \pi^i (x) .
\ee
the last terms in \eq{act:pseudoscalar} can be removed by requiring an additional relation \cite{Hong:2006ta}
\be		\la{eq:pion}
0 = \pa_z \lb \fr{z^3}{\ph^2  }  \pa_z \ls  \fr{f_0^i}{z} \rs \rb - 4 g_5^2 f_0^i .
\ee
Since the mode functions of pion satisfy the same equation, the same boundary
conditions lead to the same solutions. If we set
\be 		\la{res:equivv}		
f_0  \equiv f_0^1=f_0^2=f_0^3 ,
\ee
the following analytic solution in the chiral limit ($m_q = 0$) is allowed
with two integration constants, $N$ and $c$,
\be     \la{res:pionfunctions}
f_0 =  N z^3   \lb  I_{2/3}\left( 2 \sigma  g_5  z^3 /3 \right) -  c  \
   I_{-2/3}\left(  2  \sigma  g_5  z^3 /3   \right) \rb ,
\ee
where $ I_{\pm 2/3}$ are the modified Bessel functions.
In \cite{Kim:2009bp}, $c$ has been fixed by imposing the Dirichlet boundary condition,
$f_0 (z_{IR}) = 0$,
\be
c = \fr{I_{2/3}\left(2  \sigma  g_5 z_{IR}^3 /3 \right)}{I_{-2/3}\left( 2 \sigma  g_5 z_{IR}^3 /3  \right)} ,
\ee
and the overall normalization constant $N$ was fixed by the normalization condition
\cite{Hong:2006ta,Maru:2009ux}
\be
1 = \int_0^{z_{IR}}  dz \lb \fr{1}{2 g_5^2 z} f_0^2 + \fr{z^3}{16 \ph^2 g_5^4} \lc \pa_z
\ls \fr{f_0}{z}\rs \rc^2 \rb .
\ee
Using $1/\k^2 = N_c^2 / (4 \pi^2 R^3)$ and $1/g_5^2 = N_c^2 / (4 \pi^2 R)$
together with $N_c=3$, the numerical value of
the normalization constant reads $N= 0.3781$.

\section{Couplings of the pion interactions}

In the previous section, we have taken into account the effect of the isospin chemical potential in the isospin medium
which shifts masses of the charged mesons. We also expect the similar result for nucleons, as will be shown,  
because they have non-vanishing isospin charges. Before preforming the study on the nucleon mass splitting, in this section
we study the isospin medium effect on the couplings of the meson interactions.

From the original action \eq{2}, the terms governing the pion interactions comes from the kinetic term of the complex 
scalar field. The other terms describe the interactions of $\r$- and $a_1$-mesons which are not affected from the isospin
medium. In the isospin medium and at the rest frame, 
the kinetic term for the complex scalar can be further decomposed  into
\bea		\la{act:pionintf}
- \int d^{5}x\sqrt{G}  \ \left| D \Ph \right|^2 &=&  - \int d^{5}x \sqrt{G} \  \ph^2 \ \Tr  \lc G^{00} \ls \pa_0 e^{-i P} - i \lb V_0 , e^{-i P} \rb \rs  
\ls \pa_0 e^{i P} - i \lb V_0 , e^{i P} \rb  \rs \rp \nn
&& \qquad   + G^{mn} \ls  i  e^{-i P} l_m  -  i   r_m  e^{-i P} \rs  
\ls   -  i   l_m  e^{i P}  + i  e^{i P} r_m  \rs   \nn
&& \qquad  \lp + G^{zz} \ls \pa_z e^{-i P} + i  e^{-i P} l_z  -  i   r_z  e^{-i P}  \rs 
\ls \pa_z e^{i P} -  i   l_z  e^{i P}  + i  e^{i P} r_z  \rs \rc ,
\eea
where $V_0$, $P$, $l_M$ and $ r_M$ mean $V_0^a T^a$, $P^a T^a$, $l_M^a T^a$ and $r_M^a T^a$ respectively.
Here, the first and third terms represent the pion interactions and the second describes 
pions interacting with $\r$- and $a_1$-mesons.

From now on we concentrate only on the first term
because only it includes the isospin medium effect. 
Expanding the exponential functions, the quadratic order terms
shift the pion mass. At cubic order, the sum of the three pion interactions 
become zero automatically. At quartic order, after performing the trace, the kinetic term for the complex scalar
leads to
\bea
\d K &=&  - \fr{1}{24} \int d^{5}x\sqrt{G}  \ \ph^2 G^{00} \lb   \ls P^a \pa_0 P^a \rs^2 - P^a P^a \ \pa_0 P^b \pa_0 P^b  
-   (V_0^3)^2 \  ( P^a P^a)^2 \rp \nn
&& \qquad \qquad \qquad  \quad \lp \qquad \quad +   2 V_0^3 \e^{3bc} \lc P^a P^a (\pa_0 P^b ) P^c +  P^a (\pa_0 P^a )  P^b P^c \rc \rb .
\eea
Introducing
\be
P^{1} = \fr{1}{\sqrt{2}} \ls P^+ + P^- \rs \  , \quad P^{2} = \fr{i}{\sqrt{2}} \ls P^+ - P^- \rs  \ {\rm and } \quad P^3 = \sqrt{2} P^0,
\ee
and their Fourier mode expansions
\bea
P^0 (z,\vec{x}) &=& \int \fr{d \bar{\o}_0}{2 \pi} \ e^{i \bar{\o}_0 t}
\ P^0 (z,\bar{\o}_0) , \nn
P^{\pm} (z,\vec{x}) &=& \int \fr{d \bar{\o}_0}{2 \pi} \ e^{i \bar{\o}_{\pm} t}
\ P^{\pm} (z,\bar{\o}_{\pm}) ,
\eea
the above quartic order interactions finally reduce to the four pion interactions
in the dual field theory
\bea
\d K &=& \fr{( V^3_0) ^2 }{6}  \int d^{5}x\sqrt{G} \ \phi ^2 \ G^{00}
  \lb P^0 P^0 P^- P^+ +  P^0 P^0  P^0 P^0   \rb  \nn
 &=&  \int d^{4} x  \ls \d g_{00+-} \ \pi^0 \pi^0 \pi^+ \pi^-  + \d g_{0000}  \ \pi^0 \pi^0  \pi^0 \pi^0   \rs
\eea
where \eq{res:unitarresult} and \eq{res:modespion} are used in the last relation.
If $V^3_0=0$, these interaction terms disappear. In the isospin medium ($V^3_0 \ne 0$), 
they give rise to nontrivial effect on the four pion interactions. 
In addition, due to the relation in \eq{res:equivv} the couplings for the four pion interactions read
\bea
 \d g_{00+-}  &=& \d g_{0000}  
 =   \fr{( V^3_0) ^2  }{6}  \int d z \sqrt{G}  \ \phi ^2      \  G^{00}
 \ls  \fr{z^3}{2 \sqrt{2} \  g_5^2 \ph^2} \pa_z \ls \fr{f_0}{z} \rs   \rs^4 .
\eea
Note that the third term  in \eq{act:pionintf} is independent of the isospin medium and 
can generate all possible pion interactions, including the four pion interactions, 
which should satisfy the isospin conservation.
Let us denote such four pion interactions as $g_{ABCD} \ \pi^A \pi^B \pi^C \pi^D$
where $\{ A,B,C,D \} = \{ +,0,-\} $.
In the isospin medium, then, the couplings of four pions, $\pi^0 \pi^0 \pi^+ \pi^-$ and $\pi^0 
\pi^0  \pi^0 \pi^0 $, are shifted into
 $ g_{00+-}  + \d g_{00+-}$ and  $ g_{0000}  + \d g_{0000}$ respectively,
whereas couplings of the other four pion interactions
are not modified.  

\section{Nucleons in the isospin medium}

As shown in the previous section, the background isospin matter shifts
the masses of the charged mesons. So one can easily expect that there is also similar
mass shift in the nucleon mass because nucleons have also nonzero isospin charges.
In this section, we will investigate the nucleon mass splitting in the isospin medium.
To do so, we take into account the fermionic fluctuations on the $5$-dimensional thermal AdS background
\begin{eqnarray}		\la{act:fermion}
S &=&\int d^{5}x\sqrt{G}\left[ i \overline{\Psi }^{1} \Gamma ^{M}\nabla_{M}\Psi^{1}
+ i \overline{\Psi }^{2} \Gamma ^{M}\nabla_{M}\Psi^{2}
- m_{1}\overline{\Psi}^{1}\Psi^{1}-m_{2}\overline{\Psi }^{2}\Psi^{2} \rp \nn
&& \qquad \qquad - \lp g_Y  \ls
\overline{\Psi}^{1} \Ph \Psi^{2} + \overline{\Psi}^{2} \Ph^{+} \Psi^{1} \rs
\right] ,\label{42}
\end{eqnarray}
where $\nabla_{M}$ denotes a covariant derivative including the spin and gauge connections
\be
\na_M \Ps^{(1,2)} = \ls \pa_M - \fr{i}{4} \o_M ^{AB}\Gamma _{AB}- i V_M \rs \Ps^{(1,2)} ,
\ee
where $\Gamma^{AB}=\frac{1}{2i}\left[\Gamma^{A},\Gamma^{B}\right]$
and $V^3_t = L^3_t = R^3_t$ has been used. Here,
$(1,2)$ implies either $1$ or $2$ and $g_Y$ is the Yukawa coupling.
In the above action,
$\Psi^{1}$ and $\Psi^{2}$ transform as $\ls \frac{1}{2},0 \rs$ and $\ls 0,\frac{1}{2} \rs$
under the flavor group $SU(2)_{L}\times SU(2)_{R}$. Due to the Yukawa term,
$\Psi^{1}$ and $\Psi^{2}$ are coupled to each other and then the chiral symmetry is broken down.
This chirality can be also related to the chirality of the boundary fermion, proton and neutron. To do that,
one should take $m_1 = 5/2$ and at the same time $m_2 = - 5/2$. Then, following the AdS/CFT relation
\be
m_{(1,2)}^2=\left( \Delta - 2 \right)^{2} ,
\ee
all dual fermionic operators of $\Psi^{1}$ and $\Psi^{2}$ have the conformal dimension
of nucleon, $\D=9/2$.

If ignoring the Yukawa term, the variation of action leads to the Dirac equation
\begin{equation}
\left[i \Gamma^{M}\nabla_{M} - m_{(1,2)} \right]\Psi^{(1,2)}=0 \label{47}
\end{equation}
and the following boundary term
\begin{equation}
\lp \delta\overline{\Psi}^{(1,2)}  \Gamma ^{M}\Psi^{(1,2)} \right|^{z_{IR}}_{\e}=0  , \label{48}
\end{equation}
where $z_{IR}$ and $\e$ are the IR and UV cutoff respectively.
These are equations on the curved manifold, so it is more convenient to introduce quantities
defined on the tangent manifold in order to solve the Dirac equation.
The vielbein $e_{M}^{A}$ of the thermal AdS space is chosen as $e_{M}^{A}=\frac{1}{z} \d _{M}^{A}$,
where $A, B$ and $M,N$ are indices of the tangent and curved manifold respectively.
The non-zero components of spin connection $\omega_{M}^{AB}$ are
given by $\omega _{M }^{5A}=\frac{1}{z}\delta _{M }^{A}$. In addition, we choose the following gamma
matrix on the tangent space
\begin{equation}
\Gamma ^{t}=\left(
\begin{array}{cc}
0 & -1 \\
-1 & 0%
\end{array}%
\right) ,\quad \Gamma ^{i}=\left(
\begin{array}{cc}
0 & \sigma ^{i} \\
-\sigma ^{i} & 0%
\end{array}%
\right) ,\quad
\Gamma ^{z}=\left(
\begin{array}{cc}
-i & 0 \\
0 & i%
\end{array}%
\right)  ,  \label{49}
\end{equation}
where $\s^i$ is the Pauli matrix. If we further define the $4$-dimensional gamma matrix
$\g^{\m} = \G^{\m}$ ($\m=0,1,2,3$),
then the $4$-dimensional chiral operator  is given by $\g^5 = i \G^z$.
Using the vielbein, the above Dirac equation simply reduces to
\begin{eqnarray}		\la{eq:equationferm}
0 &=& \lb i e^M_A \G^A \left(\partial _{M} - \frac{i}{4}\omega _{M}^{AB}\Gamma _{AB}-
i  V_M \right) - m_{1} \rb \Psi^{1}
- g_Y \ph \Ps^{2} , \nn
0 &=& \lb i e^M_A \G^A \left(\partial _{M} - \frac{i}{4}\omega _{M}^{AB}\Gamma _{AB}-
i  V_M \right) - m_{2} \rb \Psi^{2}
- g_Y \ph \Ps^{1} ,\label{43}
\end{eqnarray}%
where $\ph$ is the modulus of $\Ph$ in \eq{res:scalarmodulus}.

Introducing the Fourier mode of fermion
\be
f^{(1,2)} (p,z) \ \ps^{(1,2)} (p) = \int d^4 x \ \Ps^{(1,2)} (x,z) \ e^{ i p x} ,
\ee
and the Weyl spinor representation with $\ps^{(1,2)}_L = \g^5 \ps^{(1,2)}_L$ and
$\ps^{(1,2)}_R = - \g^5 \ps^{(1,2)}_R$
\be
\ps^{(1,2)} = \ls \begin{array}{c}
\ps^{(1,2)}_L\\
\ps^{(1,2)}_R
\end{array} \rs ,
\ee
the $4$-dimensional Weyl spinors satisfy
\be
\not{\!}p \ \ps^{(1,2)}_{L,R} (p)  = \left| p \right| \ps^{(1,2)}_{R,L} (p)  ,
\ee
where the subscripts, $L$ or $R$, denote the $4$-dimensional chirality.
If one takes the
normalizable modes as $f^{1}_L$ and $f^{2}_R$ for $\Ps^{1}_L$ and $\Ps^{2}_R$
respectively, the chirality of the $SU(2)_L \times SU(2)_R$ flavor group can be associated
with the $4$-dimensional chirality.
In terms of the Weyl fermions, the equations in \eq{eq:equationferm} are reduced to
\bea
\ls \begin{array}{cc}
\pa_z - \fr{\D}{z} & - \fr{g_Y \ph}{z} \\
- \fr{g_Y \ph}{z} &  \pa_z - \fr{4- \D}{z}
\end{array} \rs
\ls \begin{array}{c}
f^{1}_L\\
f^{2}_L
\end{array} \rs
&=& - \ls \left| p \right| - V_t \rs
\ls \begin{array}{c}
f^{1}_R\\
f^{2}_R
\end{array} \rs , \la{eq:mateq1} \\
\ls \begin{array}{cc}
\pa_z - \fr{\D}{z} & \fr{g_Y \ph}{z} \\
 \fr{g_Y \ph}{z} &  \pa_z - \fr{4 - \D}{z}
\end{array} \rs
\ls \begin{array}{c}
f^{2}_R\\
f^{1}_R
\end{array} \rs
&=& \quad  \ls \left| p \right| - V_t \rs
\ls \begin{array}{c}
f^{2}_L\\
f^{1}_L
\end{array} \rs  .  \la{eq:mateq2}
\eea
These matrix equations can be further reduces to the symmetric and anti-symmetric
combinations which describe the parity-even and parity-odd excitations
under $1 \leftrightarrow 2$ and simultaneously $L \leftrightarrow R$
transformation. As a result, the lowest nucleon spectra corresponding to proton
and neutron, which are parity-even states,
are described by the lowest excitation of the symmetric combination $f^{1}_L + f^{2}_R$
together with $f^{1}_R - f^{2}_L$. On the other hand, the parity-odd states
are represented as $f^{1}_L - f^{2}_R$ and $f^{1}_R + f^{2}_L$.
In order to investigate the parity-even mass spectra, one can impose
$f^{1}_L = f^{2}_R$ and $f^{1}_R = - f^{2}_L$,
then the above two matrix equations, \eq{eq:mateq1} and \eq{eq:mateq2}, reduce to the same
matrix equation
\bea		\la{eq:even}
\ls \begin{array}{cc}
\pa_z - \fr{\D}{z} & \fr{g_Y \ph}{z} \\
\fr{g_Y \ph}{z} &  \pa_z - \fr{4- \D}{z}
\end{array} \rs
\ls \begin{array}{c}
f^{1}_L\\
f^{1}_R
\end{array} \rs
&=&
\ls \begin{array}{cc}
- \ls \left| p \right| - V_t \rs & 0 \\
0 &  \left| p \right| - V_t
\end{array} \rs
\ls \begin{array}{c}
f^{1}_R\\
f^{1}_L
\end{array} \rs  .
\eea
Similarly, imposing $f^{1}_L = - f^{2}_R$ and $f^{1}_R = f^{2}_L$
gives rise to the matrix equation for the parity-odd states
\bea 		\la{eq:odd}
\ls \begin{array}{cc}
\pa_z - \fr{\D}{z} & - \fr{g_Y \ph}{z} \\
- \fr{g_Y \ph}{z} &  \pa_z - \fr{4- \D}{z}
\end{array} \rs
\ls \begin{array}{c}
f^{1}_L\\
f^{1}_R
\end{array} \rs
&=&
\ls \begin{array}{cc}
- \ls \left| p \right| - V_t \rs & 0 \\
0 & \left| p \right| - V_t
\end{array} \rs
\ls \begin{array}{c}
f^{1}_R\\
f^{1}_L
\end{array} \rs  .
\eea
In this case, the parity-even and parity-odd states are distinguished due to the Yukawa coupling
caused by the modulus of the complex scalar field.
Furthermore, since the nucleon has an isospin charge, it
can interact with the background isospin matter.
If the $n$-th excitation mode of $f^1_{L,R}$ is denoted by $f_{L,R}^{1(n,\pm,\pm)}$
where the first and second sign imply the parity and isospin quantum number respectively,
the parity-even state satisfying \eq{eq:even}
can be further decomposed, depending on the isospin charge, into
\bea  \la{eq:parityevennucleon}
\ls \begin{array}{cc}
\pa_z - \fr{\D}{z} & \fr{g_Y \ph}{z} \\
\fr{g_Y \ph}{z} &  \pa_z - \fr{4- \D}{z}
\end{array} \rs
\ls \begin{array}{c}
f^{1(n,+,+)}_L\\
f^{1(n,+,+)}_R
\end{array} \rs
&=&
\ls \begin{array}{cc}
- \ls \left| p \right| - \fr{V_t^3}{2} \rs & 0 \\
0 &  \left| p \right| -  \fr{V_t^3}{2}
\end{array} \rs
\ls \begin{array}{c}
f^{1(n,+,+)}_R\\
f^{1(n,+,+)}_L
\end{array} \rs , \nn
\ls \begin{array}{cc}
\pa_z - \fr{\D}{z} & \fr{g_Y \ph}{z} \\
\fr{g_Y \ph}{z} &  \pa_z - \fr{4- \D}{z}
\end{array} \rs
\ls \begin{array}{c}
f^{1(n,+,-)}_L\\
f^{1(n,+,-)}_R
\end{array} \rs
&=&
\ls \begin{array}{cc}
- \ls \left| p \right| + \fr{V_t^3}{2} \rs & 0 \\
0 &  \left| p \right| + \fr{V_t^3}{2}
\end{array} \rs
\ls \begin{array}{c}
f^{1(n,+,-)}_R\\
f^{1(n,+,-)}_L
\end{array} \rs  .
\eea
On the other hand, the parity-odd states are governed by
\bea		\la{eq:parityoddnucleon}
\ls \begin{array}{cc}
\pa_z - \fr{\D}{z} & - \fr{g_Y \ph}{z} \\
- \fr{g_Y \ph}{z} &  \pa_z - \fr{4- \D}{z}
\end{array} \rs
\ls \begin{array}{c}
f^{1(n,-,+)}_L\\
f^{1(n,-,+)}_R
\end{array} \rs
&=&
\ls \begin{array}{cc}
- \ls \left| p \right| - \fr{V_t^3}{2} \rs & 0 \\
0 &  \left| p \right| -  \fr{V_t^3}{2}
\end{array} \rs
\ls \begin{array}{c}
f^{1(n,-,+)}_R\\
f^{1(n,-,+)}_L
\end{array} \rs , \nn
\ls \begin{array}{cc}
\pa_z - \fr{\D}{z} & - \fr{g_Y \ph}{z} \\
- \fr{g_Y \ph}{z} &  \pa_z - \fr{4- \D}{z}
\end{array} \rs
\ls \begin{array}{c}
f^{1(n,-,-)}_L\\
f^{1(n,-,-)}_R
\end{array} \rs
&=&
\ls \begin{array}{cc}
- \ls \left| p \right| + \fr{V_t^3}{2} \rs & 0 \\
0 &  \left| p \right| + \fr{V_t^3}{2}
\end{array} \rs
\ls \begin{array}{c}
f^{1(n,-,-)}_R\\
f^{1(n,-,-)}_L
\end{array} \rs  .
\eea
The lowest excitation modes, proton and neutron, can be represented by $f_{L,R}^{(1,+,+)}$
and $f_{L,R}^{(1,+,-)}$. Since the lowest excitations have only the parity-even states,
there is no $f_{L,R}^{(1,-,\pm)}$. For the higher resonances, they can have even and odd parity
states.

From the above matrix equations, one can easily derive two second order differential equations
for $f^{1(n,\pm,\pm)}_L$
\bea
0 &=& \lb \pa_z^2 - \fr{4 (1 - \O_{\pm})}{z} \pa_z  + \lc \fr{(5-\D - \O_{\pm}) \D}{z^2} - \fr{g_Y^2 \ph^2}{z^2}
+ \ls \left| p \right| \mp \fr{V_t^3}{2} \rs^2 \rc \rb \ f^{1(n,+,\pm)}_L , \la{eq:proton} \\
0 &=& \lb \pa_z^2 - \fr{4 (1 - \bar{\O}_{\pm})}{z} \pa_z  + \lc \fr{(5-\D - \bar{\O}_{\pm}) \D}{z^2} - \fr{g_Y^2 \ph^2}{z^2}
+ \ls \left| p \right| \mp \fr{V_t^3}{2} \rs^2 \rc \rb \ f^{1(n,-,\pm)}_L  , \la{eq:neutron}
\eea
where $\O_{\pm}$ and $\bar{\O}_{\pm}$ are given by
\be
\O_{\pm} =   \fr{4 g_Y \s z^3}{2 |p| z \mp V_t^3 z + 2 g_Y \ph} 
\quad {\rm and} \quad
\bar{\O}_{\pm} =  \fr{4 g_Y \s z^3}{2 |p| z \mp V_t^3 z - 2 g_Y \ph}  . 
\ee
The mass of nucleon is given by $p$ satisfying the following two boundary conditions
\be		\la{bcon:nucleon}
f^{1(n,\pm,\pm)}_L (0) = 0 \quad  {\rm and} \quad  f^{1(n,\pm,\pm)}_R (z_{IR}) = 0 ,
\ee
which satisfy the boundary condition in \eq{48}.
In \cite{Hong:2006ta}, the best parameters fitting the lowest nucleon masses were given by
$z_{IR} = 1/(0.205 \ {\rm GeV})$ and $g_Y=14.4$ and several nucleon masses have been investigated
for $V_t^3=0$. In the meson spectra of the hard wall model, the higher excitation modes usually have
quite different masses from the phenomenological data. In the nucleon mass the same thing happens
\cite{Hong:2006ta}.

Now, let us take into account the mass splitting of nucleons in the isospin medium.
Assuming that the nucleon mass is given by $p_0$ for $V_t^3=0$ 
in which proton, $f^{1(n,+,+)}_L$, and neutron, $f^{1(n,+,-)}_L$,
are indistinguishable. In the isospin medium with a nonzero $V_t^3$,
the proton and neutron masses are shifted due to the isospin interaction.
Since the isospin medium has a constant $V_t^3$, the boundary conditions in \eq{bcon:nucleon}
lead to $p_0 = p_P - V_t^3 /2$ for proton and
$p_0 = p_N + V_t^3 /2$ for neutron. As a result,
the proton and neutron masses in the isospin medium are given by
\bea		\la{sol:nucleonmass}
p_P &=& p_0 +  \fr{\pi^2}{\sqrt{2}}  \ls \m_P - \m_N \rs , \nn
p_N &=& p_0 -  \fr{\pi^2}{\sqrt{2}}  \ls \m_P - \m_N \rs ,
\eea
where $p_0$ generally depends on $z_{IR}$, $g_Y$, $m_q$ and $\s$.
This result shows that the proton (or neuton) mass
increases (or decreases) when the isospin chemical potential difference, $\m_P - \m_N$, grows.
As a result, the isospin interaction between nucleon and the isospin matter shifts the nucleon mass
so that like the charged meson case there exists a nucleon mass splitting proportional to the total
isospin chemical potential of the isospin matter.
Since nucleon has the half isospin charge of the charged meson
the nucleon mass shift, when $\m_P - \m_N$ is given, is exactly half of the meson mass shift.
Related solutions of \eq{eq:parityevennucleon} and \eq{eq:parityoddnucleon},
there is a remarkable point. After inserting the mass relations in \eq{sol:nucleonmass} back into
the equations of motion, \eq{eq:parityevennucleon} and \eq{eq:parityoddnucleon}, one can easily find
\be		\la{res:isospinindependence}
f^{1(m,\pm,+)}_L = f^{1(m,\pm,-)}_L \quad {\rm and} \quad f^{1(m,\pm,+)}_R = f^{1(m,\pm,-)}_R ,
\ee
which implies that in spite of the nucleon mass shiftings the Fourier mode functions of nucleons are
not changed. 
Another interesting limit is the chiral phase transition point where the chiral condensate disappears
and $\ph$ simply reduces to $\ph = m_q z$. If the nucleon mass is given by $m_0$ for
$m_q = V_t^3=0$, the proton mass from \eq{eq:proton} yields
\be
p_P = \sqrt{m_0^2 + g_Y^2 m_q^2 }  + \fr{\pi^2}{\sqrt{2}}  \ls \m_P - \m_N \rs ,
\ee
whereas the neutron mass from \eq{eq:neutron} is given by
\be
p_N = \sqrt{m_0^2 + g_Y^2 m_q^2 }  - \fr{\pi^2}{\sqrt{2}} \ls \m_P - \m_N \rs .
\ee
These results, as expected, show that proton becomes more massive in the
isospin medium with $\m_P \gg \m_N$ so that in this case it is easier to
create neutron than proton.

\section{Pion-nucleons coupling}

From  the fermion action \eq{act:fermion}, the interaction term between nucleons and pion at cubic order
is described by
\bea
S_{int} &=& i \int d^4 x \int_0^{z_{IR}} \fr{dz}{z^4}
\lb \ls {\ps^1_R}^+  l_z \ps^{1}_L -  {\ps^{1}_L}^+  l_z \ps^{1}_R
+ {\ps^{2}_R}^+  r_z \ps^{2}_L -  {\ps^{2}_L}^+  r_z \ps^{2}_R \rs \rp \nn
&& \lp \qquad \qquad \qquad \qquad +  g_Y  \ph
\ls  {\ps^{1}_R}^+  P   \ps^{2}_L  +  {\ps^{1}_L}^+  P   \ps^{2}_R
-  {\ps^{2}_R}^+  P   \ps^{1}_L  -  {\ps^{2}_L}^+  P   \ps^{1}_R   \rs  \rb,
\eea
where $l_z$ and $r_z$ are left and right gauge fluctuations of $SU(2)_L \times SU(2)_R$ in the $z$-direction.
Using \eq{def:vectoraxial}, $v_z$ is decoupled in the unitary gauge as mentioned before.
Introducing further the following notations
\bea
\ls \begin{array}{c} \ps^{1}_L \\ \ps^{2}_L  \end{array} \rs
= \ls \begin{array}{c} f_L^{1(n,\pm,\pm)} \ps_L^{(n,\pm,\pm)}
\\ f_L^{2(n,\pm,\pm)} \ps_L^{(n,\pm,\pm)}   \end{array} \rs \quad {\rm and} \quad
\ls \begin{array}{c} \ps^{1}_R \\ \ps^{2}_R  \end{array} \rs
= \ls \begin{array}{c} f_R^{1(n,\pm,\pm)} \ps_R^{(n,\pm,\pm)}
\\ f_R^{2(n,\pm,\pm)} \ps_R^{(n,\pm,\pm)}   \end{array} \rs ,
\eea
the parity-even and parity-odd modes satisfy
\bea		\la{rel:parityrel}
f_L^{1(n,+,\pm)} = f_R^{2(n,+,\pm)} , \quad
f_R^{1(n,+,\pm)} = - f_L^{2(n,+,\pm)} , \nn
f_L^{1(n,-,\pm)} = - f_R^{2(n,-,\pm)} , \quad
f_R^{1(n,-,\pm)} = f_L^{2(n,-,\pm)} .
\eea
Since the lowest nucleon resonances, proton and neutron denoted by $\Ps^{(1,+,+)} $ and $\Ps^{(1,+,-)}$ respectively, have the positive parity, the interaction between these lowest nucleons and pion leads to
\be
S_{int} = i \sqrt{2} \int d^4 x \ g_{(1,+,\pm;n,+,\pm)}  \ \bar{\Ps}^{(1,+,\pm)} \g^5 \pi \Ps^{(1,+,\pm)} .
\ee
In this case,  the nucleons-pion coupling $g_{(1,\pm,\pm;1,\pm,\pm)} $ is given by
\bea
g_{(1,+,\pm;1,+,\pm)} &=&  \int_0^{z_{IR}} \fr{dz}{z^4}  \lb -  \fr{f_0}{2} \ls  {f^{1(1,+,\pm)}_L}^* f^{1(1,+,\pm)}_R +  {f^{1(1,+,\pm)}_R}^* f^{1(1,+,\pm)}_L \rs \rp \nn
&& \lp + \fr{g_Y z^3}{4  g_5^2 \ph} \pa_z \ls \fr{f_0}{z} \rs
\ls {f^{1(1,+,\pm)}_L}^* f^{1(1,+,\pm)}_L +  {f^{1(1,+,\pm)}_R}^* f^{1(1,+,\pm)}_R  \rs \rb ,
\eea
where $f_0$ means the Fourier mode function of pions in \eq{res:pionfunctions}.
and  $\Ps^{(1,+,\pm)}$ implies the $4$-dimensional Dirac fermion
\bea
\Ps^{(1,+,\pm)} = \ls \begin{array}{c}
\ps_L^{(1,+,\pm)}\\
\ps_R ^{(1,+,\pm)}
\end{array} \rs .
\eea
Note that the interaction terms can be further decomposed according to the isospin charges of nucleons.

If the pion, $\pi = \pi^a t^a$, is decomposed following its isospin charge into
\be
\pi^{1} = \fr{1}{\sqrt{2}} \ls \pi^+ + \pi^- \rs \  , \quad \pi^{2} = \fr{i}{\sqrt{2}} \ls \pi^+ - \pi^- \rs  \ {\rm and } \quad \pi^3 = \sqrt{2} \pi^0,
\ee
the allowed interaction terms in the isospin medium become
\bea		\la{eq:couplingproneut}
S_{int} &=& i  g_{\pi NN}  \int d^4 x   \ \ls \bar{P} \g^5 \pi^0 P
-   \bar{N} \g^5 \pi^0 N    +   \bar{P} \g^5 \pi^+ N  +   \bar{N} \g^5 \pi^- P \rs   ,
\eea
where proton and neutron are denoted by $P=\Ps^{(1,+,+)} $ and $N=\Ps^{(1,+,-)} $
and the pion-nucleon coupling $g_{\pi NN}$ satisfies
\be		\la{res:degenerate}
g_{\pi NN} \equiv  g_{(1,+,+;1,+,+)} = g_{(1,+,-;1,+,-)} = g_{(1,+,+;1,+,-)}
= g_{(1,+,-;1,+,+)} \ .
\ee
This result implies that the same mode function of nucleons
in the isospin medium
makes the pion-nucleon coupling independent of the isospin charges and
the total isospin chemical potential, as mentioned before.
As a result, we find that there is no pion-nucleon coupling splitting in the isospin medium 
while the nucleon and meson masses are splitted due to the isospin interaction.

In \cite{Lee:2013oya} the following parameters, $z_{IR}= 1/( 0.3227 \ {\rm GeV})$,  
$m_q = 0$ and $\s =(0.304 \ {\rm GeV} )^3$, have been used to obtain good lowest meson spectra.
Using these, the mass of proton and neutron, $p_0 = 0.9399 \ {\rm GeV}$, for $\m_P - \m_N=0$
can be obtained by taking $g_Y = 4.762$. Furthermore,  the following normalization condition
\be
1 = \int_0^{z_{IR}} \fr{dz}{z^4}  \ls \left| f^{1(m,\pm,\pm)}_L \right| ^2 +
\left| f^{1(m,\pm,\pm)}_R \right| ^2\rs ,
\ee
gives rise to $g_{\pi NN}=11.1675$, which
is almost consistent with the experimental data $g_{\pi NN}=13.1$ \cite{Kim:2009bp} like the meson case.
The analysis used in this paper can be easily generalized to the higher resonance cases.

\section{Discussion}

In this paper, we have investigated the nucleon spectra and pion-nucleon coupling
in the isospin medium which can provide a good
playground to study and understand the medium effect due to
its simplicity and analyticity. Here we considered a hard wall model with a negative cosmological constant and constant gauge potentials corresponding to the isospin chemical potentials.
In general, since the equations of motion do not depend on the gauge potential
but on the gauge field strength, the bulk constant gauge potential
does not modify the background geometry. However, since bulk fermions are coupled to
the gauge potential, some physical quantities associated with fermion can be affected
by the constant gauge potentials. This is the story of the gravity theory. Following the
AdS/CFT correspondence, the $5$-dimensional gravity theory can be reinterpreted
as physics of the $4$-dimensional dual gauge theory.

In the dual field theory, the constant gauge potentials are mapped to the isospin chemical potentials
and the bulk massive fermions which have a mass, either $5/2$ or $-5/2$,
are dual to nucleons with a conformal dimension $9/2$. If there is no nontrivial gauge
potential, proton and neutron are indistinguishable, so they have the same physical
properties like mass spectrum and pion-nucleon coupling.
In the isospin medium proton and neutron interact with the isospin matter
in a different way due to their different isospin charges so that the nucleon mass splitting
occurs. More precisely, the nucleon mass splitting
is exactly half of the meson mass splitting because the nucleon isospin charge
is half of charged mesons.
As expected, the masses of mesons and nucleons
increase or decrease linearly as $\m_P - \m_N$ increases.

We also studied the couplings of pion interactions in the isospin medium.
The isospin medium does not affect on the couplings of the three pion interactions. 
However, we found that the nontrivial isospin chemical potential of the isospin medium modifies
the coupling only for $\pi^0 \pi^0 \pi^+ \pi^-$ and $\pi^0 
\pi^0  \pi^0 \pi^0 $ of the four pion interactions.
For the pion-nucleon couplings, in spite of the mass shifts of nucleons and mesons,
there is no effect from the isospin medium because the mode functions
are independent of the background isospin chemical potential.

It would be interesting to apply holographic techniques studied in this paper
to the nuclear medium for understanding the more real physical systems
like the nuclear matter and neutron star.
Although the hard wall model leads to almost consistent lowest mass spectra
with the phenomenological data, the high resonance spectra are quite different from the real data.
So, it would be interesting to improve further the hard wall model in order to explain the mass spectra of high excitations
well.

\vspace{1cm}

{\bf Acknowledgements}

We thank J. Erlich for valuable discussion.
Sh.M. thanks CQUeST members for hospitality during his visit to this center.
C.P. thanks APCTP for hospitality during his visit.
B.-H.L. and Sh.M.  were supported by the National Research Foundation of Korea (NRF) grant funded by
the Korea government (MEST) through the Center for Quantum Spacetime (CQUeST) of Sogang
University with grant number 2005-0049409.
Sh.M. was supported by the Science Development Foundation under the President of Republic of Azerbaijan Grant № EIF-Mob-2-2013-4(10)-13/02/1.  C.P. was supported by the WCU grant no. R32-10130, the Research fund no. 1-2008-2935-001-2 by Ewha Womans University and  also supported by Basic Science Research Program through the National Research Foundation of
Korea (NRF) funded by the Ministry of Education (Grant No. NRF-2013R1A1A2A10057490).

\vspace{1cm}


\end{document}